\documentclass[prl,twocolumn,showpacs,flushleft,amsmath,amssymb,nofootinbib]{revtex4}
\usepackage{subfig}
\usepackage{graphicx}
\usepackage{dcolumn} 
\usepackage{bm}      
\usepackage[active]{srcltx} 
\usepackage{hyperref} 
\usepackage{amsmath}
\usepackage{amssymb}
\usepackage{stmaryrd}
\usepackage{amsfonts}
\usepackage{mathrsfs}
\usepackage{txfonts}

\usepackage{bm}

\newcommand{\Ref}[1]{(\ref{#1})}

\newcommand{\Z}{\mathbb{Z}}
\newcommand{\R}{\mathbb{R}}

\newcommand{\half}{\frac{1}{2}}

\def\be{\begin{eqnarray}}
\def\ee{\end{eqnarray}}

\newcommand{\sa}{\mathscr{A}}
\newcommand{\sz}{\mathscr{Z}}

\renewcommand{\a}{\alpha}
\renewcommand{\b}{\beta}
\newcommand{\g}{\gamma}

\newcommand{\eps}{\varepsilon}

\renewcommand{\l}{\lambda}

\newcommand{\rmd}{\mathrm d}

\newcommand{\lt}{\left}
\newcommand{\rt}{\right}

\newcommand{\tr}{\mathrm{tr}}

\begin{document}

\sloppy

\title{\bf Black Hole Entropy in Loop Quantum Gravity, Analytic Continuation, and Dual Holography}

\author{Muxin Han }

\affiliation{Centre de Physique Th\'eorique
, CNRS UMR7332, Aix-Marseille Universit\'e and Universit\'e de Toulon, 13288 Marseille, France}


\begin{abstract}

A new approach to black hole thermodynamics is proposed in Loop Quantum Gravity (LQG), by defining a new black hole partition function, followed by analytic continuations of Barbero-Immirzi parameter to $\g\in i\R$ and Chern-Simons level to $k\in i\R$. The analytic continued partition function has remarkable features: The black hole entropy $S=A/4\ell_P^2$ is reproduced correctly for infinitely many $\g= i\eta$, at least for $\eta\in\Z\setminus\{0\}$. The near-horizon Unruh temperature emerges as the pole of partition function. Interestingly, by analytic continuation the partition function can have a dual statistical interpretation corresponding to a dual quantum theory of $\g\in i\Z$. The dual quantum theory implies a semiclassical area spectrum for $\g\in i\Z$. It also implies that at a given near horizon (quantum) geometry, the number of quantum states inside horizon is bounded by a holographic degeneracy $d= e^{A/4\ell_P}$, which produces the Bekenstein bound from LQG. The result in \cite{FGNP} also receives a justification here.

\end{abstract}


\maketitle

It is well-known that black hole, as a system arise from General Relativity (GR), has remarkable thermodynamical properties \cite{BH}. In particular, black hole has an entropy proportional to its area by $S=A/4\ell_P^2$. The black hole entropy results in important ramifications such as the Bekenstein's entropy bound, and the covariant Bousso's bound \cite{bound}, which conjectures that the number of microstates inside a (spatial) region is bounded by $e^{A/4\ell_P^2}$ where $A$ is the area surrounding the region. The conjecture leads to the holographic principle for quantum gravity \cite{holo}. 

The statistical origin of black hole entropy needs to be explained by quantum gravity. In this paper we propose a new approach to black hole entropy in Loop Quantum Gravity (LQG) \cite{rev}. There has been a long history of computing black hole entropy from LQG (e.g. \cite{BHLQG,ENP,revisit}). The resulting black hole entropy has had a famous dependence of Barbero-Immirzi parameter $\g\in\R$. Reproducing $S=A/4\ell_P^2$ relies on fine-tuning $\g$ to a single critical value $\g_0$. The situation is improved by the recent progress \cite{GP,FGP}, where an area-energy relation $E=\frac{A}{8\pi\ell}$ allows to equivalently formulate black hole as a (grand) canonical ensemble. However it still has not been clear yet how exactly $A/4\ell_P^2$ emerges as black hole entropy from LQG framework.  

In this work, a new grand canonical partition function $\sz$ is proposed for LQG black hole. Then we analytic continue the partition function to purely imaginary Barbero-Immirzi parameter $\g\in i\R$ (up to a small real part). Correspondingly, the Chern-Simons level is complexified $k\in i\R$, motivated by a relation between $k$ and $\g$ in isolated horizon context \cite{ENP}. Motivated by \cite{FGNP}, we take the viewpoint that an object of LQG with complex $\g$ is defined by the corresponding object from well-defined quantization with real $\g$, followed by an analytic continuation of $\g$ to complex plane. Interestingly, the analytic continuation results in the following remarkable features: 

\begin{itemize}

\item The analytic continued black hole partition function $\sz$ reproduces correctly the entropy $S=A/4\ell_P^2$ as the leading contribution, supplemented by quantum and UV corrections.

\item The derivation works at least for $\g\simeq i\eta$ ($\eta\in\Z\setminus\{0\}$) up to small real part. There are infinitely many allowed purely imaginary $\g$, all resulting in $S=A/4\ell_P^2$. The case of Ashtekar's variables \cite{ash} is included as $\g=\pm i$. Generalization to noninteger $\eta$ may rely on a technical assumption of analytic continuation. 

\item The Unruh temperature $\b_U=\frac{2\pi\ell}{\ell_P^2}$ (of near horizon observer with distance $\ell$) appears as a pole in the analytic continued partition function. The naturality of $\b_U$ is also suggested by \cite{P} from a different point of view.

\item Close to Unruh temperature, $\sz$ can have a \emph{dual} interpretation as a statistical system, corresponding to a \emph{dual quantum theory} associated with $\g=i\eta$. The resulting dual quantum theory has a (semiclassical) area spectrum $A=8\pi|\eta|\ell_P^2\sum_ls_l\ (s_l\in\R_+)$ up to a specific rescaling.

\item More importantly, in the dual quantum theory, at a given near horizon (quantum) geometry, the number of quantum states inside horizon is bounded by the degeneracy $d\simeq e^{A/4\ell_P}$. Such a holographic degeneracy produces the Bekenstein bound from LQG. The assumption of holographic degeneracy in \cite{GNP} also receives a justification here.

\end{itemize}

On the other hand, the positivity of black hole energy spectrum, the analyticity (holomorphicity), and the existence of dual statistical interpretation of $\sz$, suggests a specific 1st order quantum correction to the classical energy-area relation proposed in \cite{GP}. The correction may come from the radiative correction from LQG \cite{aldo}.

Black hole in LQG is described in terms of an SU(2) Chern-Simons theory with level $k$ \cite{ENP}
\be
S_{CS}[\sa]=\frac{k}{4\pi}\int_{H}\tr\lt(\sa\wedge\rmd\sa+\frac{2}{3}\sa\wedge\sa\wedge\sa\rt)
\ee
where $H$ is the black hole horizon with spatial area $A$. The Chern-Simons level $k$ will be complexified to $k\in i\R$ as analytic continuing $\g\in i\R$. 


The near-horizon quantum geometry of black hole are described by $N$ punctures on spatial section of $H$ from Wilson lines in Chern-Simons theory, with a set of spins/areas $\{j_l\}_{l=1}^N$ \cite{ENP,bianchi}. Given $\{j_l\}_{l=1}^N$, A quantum state inside horizon is a Chern-Simons state on $S^2$ with $N$ punctures colored by $\{j_l\}_{l=1}^N$. The Hilbert space has the dimension given by the famous Verlinde formula \cite{verlinde,BT} ($d_l=2j_l+1$):
\be
\dim_{k}\lt(\vec{j}\rt)=\frac{2}{k+2}\sum^{k+1}_{d=1}\sin^2\lt(\frac{\pi d}{k+2}\rt)\prod_{l=1}^N\frac{\sin\lt(\frac{\pi d d_l}{k+2}\rt)}{\sin\lt(\frac{\pi d}{k+2}\rt)}\label{verlinde}
\ee
which is the degeneracy the black hole microstates at a given near-horizon geometry. The LQG approach of black hole entropy has been based on the Verlinde formula, which has led to the well-known $\g$-dependence \cite{revisit}. Recently there has been an interesting observation from \cite{FGNP}: if $\dim_k(\vec{j})$ are analytic continued to $j_l=is_l-1/2$, its asymptotic behavior as $s_l$ large gives $e^{A/4\ell_P}$, in terms of a conjectured LQG area spectrum when $\g=\pm i$. The result motivates that the $A/4\ell_P$-law may naturally come from a quantum theory with complex Ashtekar connection with purely imaginary $\g$. Such viewpoint motivates the work here and has been adopted in several recently works \cite{igamma}. However such an interesting result in \cite{FGNP} is mysterious and has to be justified. When $j$ is complexified, the Verlinde formula loses the meaning as a Hilbert space dimension. It has not been clear yet if the result in \cite{FGNP} counts the quantum states of any system. Such an issue in \cite{FGNP} will be justified in the following analysis.

Let's consider a quantum black hole horizon described by a gas of $N$ punctures. Motivated by \cite{GP,GNP}, a canonical partition function is defined by summing over spin configurations, with a degeneracy factor given by $\dim_k(\vec{j})$:
\be
Z_N=\frac{1}{N!}\sum_{j_1\cdots j_N=\half}^{k/2}\dim_k\lt(\vec{j}\rt)\ e^{-\b E\lt(\vec{j}\rt)}\label{ZN}
\ee
The grand canonical partition function is defined by $\sz=\sum_NZ_Ne^{\mu N}$. Here $\dim_k\lt(\vec{j}\rt)$ is a faithful counting of degenerate states with a given set of $\{j_l\}_l$. $1/N!$ is a Gibbs factor of indistinguishable punctures. The Hamiltonian is defined by 
\be
E=\g\frac{\ell_P^2}{\ell}\sum_{l=1}^N\lt[j_l+\half+f(\g,k)\rt].
\ee
In the semiclassical large-$j$ regime, the energy spectrum proposed here is consistent with the LQG area $A=8\pi\g\ell_P^2\sum_{l=1}^N \sqrt{j_l(j_l+1)}$ and the classical energy-area relation $E=\frac{A}{8\pi\ell}$ of near-horizon observer \cite{GNP,GP,FGP}. $\ell$ is the small proper distance from the horizon. $f(\g,k)$ stands for a possible quantum deviation from the classical area-energy relation. It has to be a holomorphic function in order to perform analytic continuation. It has to be real as $\g,k\in\R$ for a Hermitian Hamiltonian. Our analysis will fix $f(\g,k)$ to the following form:
\be
f(\g,k)=\frac{1}{2\pi\g}\lt[m\log k+\log\a_m(\g)\rt] \label{f}
\ee 
with parameters $m\geq 0$ and $\a_m(\g)>0$ satisfying certain condition. The $\log k$ term may relates to the self-energy from spinfoam amplitude \cite{aldo}.

Here we have analytic continued $Z_N$ to complex $\g$-plane, and set $\g=-i\eta$, where $\eta=\eta_0-i\eps$ ($\eps$ small) with $\eta_0\in \Z\setminus\{0\}$. Without loss of generality, we set $\eta_0>0$ in the main content. Our following analysis is symmetric under $\eta\to-\eta$.

The local temperature of the near-horizon observer is the Unruh temperature $\b_U=\frac{2\pi\ell}{\ell_P^2}$. The range of sum $\sum_j$ in Eq.\Ref{ZN} is from $\half$ to $\frac{k}{2}$, i.e. the integrable representations in $\mathrm{SU(2)}_k$ affine Lie algebra \cite{seiberg}. 

Now the summand in $Z_N$ becomes oscillatory, which would make $Z_N$ lose the interpretation as a statistical partition function. However the following procedure leads to a ``dual statistical system'', which does interpret $Z_N$ as a statiscal partition function. Insert the Verlinde formula and sum over $j_l$,
\be
Z_N=c_{N,k}\sum_{d=1}^{k+1}\sin^{2-N}\lt(\frac{\pi d}{k+2}\rt)\prod_{l=1}^N\sum_{d_l=2}^{k+1}\lt[e^{i\Delta_d^+ \frac{d_l}{2}}-e^{i\Delta_d^- \frac{d_l}{2}}\rt],
\ee
where $d_l=2j_l+1$ and
\be
\Delta_d^\pm=\eta\b\frac{\ell_P^2}{\ell}\pm \frac{2\pi d}{k+2},\ \ \ c_{N,k}=\frac{(-i)^N}{N!}\frac{2^{1-N}}{k+2}e^{Nf(-i\eta ,k)i\eta\b\frac{\ell_P^2}{\ell}}.
\ee
The sum $\sum_{d_l=2}^{k+1}$ can be performed easily. Then $Z_N$ reads
\be
c_{N,k}\sum_{d=1}^{k+1}\sin^{2-N}\lt(\frac{\pi d}{k+2}\rt)\lt[\frac{e^{i\Delta_d^+}\lt(e^{\frac{i k}{2}\Delta_d^+}-1\rt)}{e^{\frac{i}{2}\Delta_d^+}-1}
-\frac{e^{i\Delta_d^-}\lt(e^{\frac{i k}{2}\Delta_d^-}-1\rt)}{e^{\frac{i}{2}\Delta_d^-}-1}\rt]^N.\nonumber
\ee

Now we complexify the Chern-Simons level $k=i\l-2$ in the partition function, where $\l\in\R_+$ is large but finite. There is an obvious difficulty that $k$ appears as the upper bound of the sum $\sum_{d=1}^{k+1}$. However no one prevents us to firstly make the replacement $k=i\l-2$ for $k$ appearing inside the summand. After replacement $Z_N$ reads
\be
c_{N,\l}\sum_{d=1}^{k+1}\sin^{2-N}\lt(\frac{\pi d}{i\l}\rt)\lt[\frac{e^{i\Delta_d^+}\lt(e^{\frac{-\l-2i}{2}\Delta_d^+}-1\rt)}{e^{\frac{i}{2}\Delta_d^+}-1}
-\frac{e^{i\Delta_d^-}\lt(e^{\frac{-\l-2i}{2}\Delta_d^-}-1\rt)}{e^{\frac{i}{2}\Delta_d^-}-1}\rt]^N.\nonumber
\ee
$k$ appearing at $\sum_{d=1}^{k+1}$ is temporarily kept unchanged. One should firstly perform the sum then analytic continue $k$. Here $\Delta_d^\pm$ and $c_{N,\l}$ read
\be
\Delta_d^\pm=\eta\b\frac{\ell_P^2}{\ell}\pm \frac{2\pi d}{i\l},\ \ \ c_{N,\l}=\frac{(-i)^N}{N!}\frac{2^{1-N}}{i\l}e^{Nf(-i\eta ,i\l)i\eta\b\frac{\ell_P^2}{\ell}}.
\ee
The partition function has a series of nontrivial poles at 
\be
\Delta_d^\pm = 4\pi q_\pm,\ \ \ (q\in\Z,q\neq 0)
\ee
As long as $q\neq0$, the residue in each factor of summand at the pole is nonzero, thanks to the complexification of $k$. 

We firstly consider the case $\eta_0$ is an odd integer, i.e. $\eta=2 q-1-i\frac{ x}{\l}$ ($q,x\in\Z_+,x>0,x\ll\l$) with small imaginary part, it picks the $k+2-x$ term (close to the top of the sum) outside the sum $\sum_{d=1}^{k+1}$, i.e. we write the sum in $Z_N$ as
\be
\lt[\frac{e^{i\Delta_{k+2-x}^+}\lt(e^{\frac{-\l-2i}{2}\Delta_{k+2-x}^+}-1\rt)}{e^{\frac{i}{2}\Delta_{k+2-x}^+}-1}
-\frac{e^{i\Delta_{k+2-x}^-}\lt(e^{\frac{-\l-2i}{2}\Delta_{k+2-x}^-}-1\rt)}{e^{\frac{i}{2}\Delta_{k+2-x}^-}-1}\rt]^N+\sum_{d\neq k+2-x}^{k+1} \cdots\nonumber
\ee
The $k+2-x$ term outside the sum can be analytic continued to $k=i\l-2$ without difficulty. Then the Unruh temperature $\b_U=\frac{2\pi\ell}{\ell_P^2}$ appears as the pole of this term 
\be
0=\Delta^+_{i\l-x}-4\pi q=\lt(2 q-1-i\frac{ x}{\l}\rt)\frac{\ell_P^2}{\ell}\lt(\b-\b_U\rt)
\ee 
The residue of the pole within the factor is $2i$ approximately, as we ignore the exponentially decaying $e^{-\l 4\pi q}$.

Such a pole can never appear from the rest of terms in $\sum_{d\neq k+2-x}^{k+1}$, it also doesn't coincide with the pole given by $\Delta_{i\l-x}^-$. Indeed, if we pick out the $d$ term and analytic continue in the same way as above, close to $\b_U$
\be
\Delta_d^{\pm}=\lt(\Delta_{i\l-x}^+-4\pi q\rt)+\frac{2\pi (x\pm d-i\l)}{i\l}+4\pi q
\ee 
If $\Delta_d^{\pm}= 4\pi m$ with $m\in\Z$ when $\Delta_{i\l-x}^+=4\pi q$, $\frac{(x\pm d-i\l)}{i\l}$ would be an even number, which implies $d=\pm (2m+1)i\l\mp x$ after complexification. It can only happen in the $\Delta^+_d$ case with $d=i\l-x$ since originally $0<d<k+2$. However it can nevertheless happen that $\Delta_d^{\pm}=4\pi\Z+o(\frac{1}{\l})$ at $\b_U$, e.g. modulo $4\pi\Z$, $\Delta_{i\l-x}^-(\b_U)=\frac{2x}{i\l}-4\pi$ and $\Delta_{i\l-x+1}^+(\b_U)=\frac{1}{i\l}$, i.e. other terms with $d\neq k+2-x$ can have contribution of $o(\l)$. 

The next task is to show the sum $\sum_{d\neq k+2-x}^{k+1}$ is negligible if $\b$ is sufficiently close to $\b_U$. We may estimate the sum by an integral up to $o(1/k)$ i.e. we write the sum to be
\be
(k+2)\lt[\int_{\frac{1}{k+2}}^{\frac{k+2-x-1}{k+2}}\rmd \lt(\frac{d}{k+2}\rt)\cdots+\int^{\frac{k+1}{k+2}}_{\frac{k+2-x+1}{k+2}}\rmd \lt(\frac{d}{k+2}\rt)\cdots\rt]
\ee
Analytic continuation $k=i\l-2$ corresponds a rotation of integration contour ($\xi=d/\l$):
\be
\l\lt[\int_{\frac{1}{\l}}^{\frac{i\l-x-1}{\l}}\rmd \xi\cdots+\int^{\frac{i\l-1}{\l}}_{\frac{i\l-x+1}{\l}}\rmd \xi\cdots\rt]\label{estimate}
\ee
where the integrand reads 
\be
\frac{1}{\sin^{N-2}\lt(-i\pi\xi\rt)}\lt[\frac{e^{i\Delta_\xi^+}\lt(e^{\frac{-\l-2i}{2}\Delta_\xi^+}-1\rt)}{e^{\frac{i}{2}\Delta_\xi^+}-1}
-\frac{e^{i\Delta_\xi^-}\lt(e^{\frac{-\l-2i}{2}\Delta_\xi^-}-1\rt)}{e^{\frac{i}{2}\Delta_\xi^-}-1}\rt]^N
\ee
where $\Delta_\xi^\pm=\eta\b\frac{\ell_P^2}{\ell}\pm {2\pi(-i) \xi}$. By above discussion, when $\b$ close to $\b_U$, the integrand has a $N$-th order pole at $\xi=\frac{i\l-x}{\l}$. When $N>2$ it has additional ($N-2$)-th order pole at $\xi=0,i$. However all the poles have a $1/\l$-distance away from the integration contour. Since the pole $\xi=\frac{i\l-x}{\l}$ is close to $\xi=i$, the integral Eq.\Ref{estimate} grows as $\l^{2N-2}$, which is also the leading behavior of the sum $\sum_{d\neq k+2-x}^{k+1}$ after analytic continuation. On the other hand, the $d=k+2-x$ term outside the sum is of the order ${\l^{N-2}}{\delta_\b^{-N}}$. Therefore, when we are inside the regime that $\delta_\b=\eta\frac{{\ell_P^2}}{\ell}(\b-\b_U)\ll \frac{1}{\l}$, the contribution from $\sum_{d\neq k+2-x}^{k+1}$ is negligible for all $N$.

The partition function is simplified dramatically after the approximation. As $\l\gg1$
\be
Z_N\simeq \frac{1}{N!}\lt[\frac{2\pi^2 x^2}{(i\l)^3}\rt]i^Ne^{Nf(-i\eta,i\l)2\pi i\eta}\lt(\frac{\l}{\pi x}\rt)^N\lt[\frac{{\ell}}{\eta{\ell_P^2}\lt(\b-\b_U\rt)}\rt]^N.\label{star}
\ee 

If $\eta_0$ is an even integer, i.e. $\eta=2 q +i\frac{ x}{\l}$ ($q\in\Z_+,x>0,x\ll\l$), it picks up the poles close to the bottom of the sum $\sum_{d=1}^{k+1}$, i.e. the term with $d=x\in\Z_+$ with 
\be
0=\Delta^+_{x}-4\pi q=\lt(2 q+i\frac{ x}{\l}\rt)\frac{\ell_P^2}{\ell}\lt(\b-\b_U\rt)
\ee 
The estimate can be carried out in the same way as above, by replacement $x\to i\l-x$. A similar result holds, i.e. as $\eta=2 q +i\frac{ x}{\l}$, the term with $d=x$ is picked up as the leading contribution, as long as $\delta_\b\ll \frac{1}{\l}$. The resulting partition function is exactly the same as Eq.\Ref{star}.

The derivation with integer $\eta_0$ works because it is allowed to pick up terms at the top or bottom in the sum $\sum_{d=1}^{k+1}$ for analytic continuation of $k$. It may or may not work for terms in the middle. e.g. If we assume picking up $d=\frac{1}{2}(k+1)$ is allowed, the above derivation generalizes to noninteger $\eta_0$. However $\frac{1}{2}(k+1)$ may not always a integer for all $k$, the term $d=\frac{1}{2}(k+1)$ may or may not appear in the sum. So generalization to noninteger $\eta_0$ may rely on nontrivial assumptions.

In Eq.\Ref{star}, $\lt[\frac{2\pi^2 x^2}{(i\l)^3}\rt]$ only contributes the logarithmic correction in grand potential $\log\sz$. The rest part in $Z_N$ has to be real and positive in order to have a dual statistical interpretation. Thus $e^{f(-i\eta,i\l)2\pi i\eta}=\chi(-i\eta,i\l)$, where both $f$ and $\chi$ are holomorphic in $\l,\eta$ and $\chi(-i\eta,i\l)\in i\R_-$. As $f,\chi$ are holomorphic, this equation holds on the whole complex plane, which implies $e^{-f(\g,k)2\pi \g}=\chi(\g,k)$. $f(\g,k)$ is real for a Hermitian Hamiltonian $E$, which implies $\chi(\g,k)\in \R_+$. Expand $\chi$ into power series $\chi(\g,k)=\sum_{m} k^{m}\a_m(\g)$, and keep only the leading term as $k$ large. If the leading order would be of $o(k^{m>0})$, it would give $f(\g,k)=\frac{-m}{2\pi\g}\log k $ as the leading order, which would produce negative $E$ for small spins. A positive definite energy spectrum implies the leading order of $\chi(\g,k)$ is $k^{-m}\a_m(\g)$ with $m\geq0$. $\a_m(\g)$ should satisfy $\a_m(\g)\in\R_+$ and $i^{-m+1}\a_m(-i\eta)\in\R_+$. So we fix $f(\g,k)$ to the form in Eq.\Ref{f}.

Here we allow the creation and annihilation of the punctures on the horizon (or a sum over graphs in LQG terminology). We define a grand canonical partition function $\sz=\sum_NZ_Ne^{\mu N}$ where $\mu$ is a postulated chemical potential.
\be
\log\sz\simeq 
\frac{\l|\chi|}{\pi x}e^\mu\frac{{\ell}}{\eta{\ell_P^2}\lt(\b-\b_U\rt)}-3\log \l.
\ee

The leading contribution to mean energy $U=-\partial_\b\log\sz$ can be computed straightforwardly as $k$ being large:
\be
U[\b_-]\simeq \frac{\l|\chi|}{\pi x}e^\mu \frac{ {\ell}}{\eta{\ell_P^2}\lt(\b-\b_U\rt)^2}\lt[1+o(\l^{-1})\rt]
\ee
which relates the horizon area by the classical relation $U=\frac{A}{8\pi\ell}$. If $m=0$ in Eq.\Ref{f} then $\chi\sim o(1)$, we obtain the relation $\eta\frac{{\ell_P^2}}{\ell}(\b-\b_U)\equiv\delta_\b\propto \sqrt{\l{\ell_P^2}/{A}}$. If $m=1$ then $\chi\sim o(1/\l)$ and $\delta_\b\propto \sqrt{{\ell_P^2}/{A}}$. $\delta_\b$ becomes finer as $m$ increase.

The entropy from the grand canonical ensemble is given by $S=\b U+\log\sz$. The leading contribution of entropy is given by $\b U$ because $\log\sz\sim\delta_\b^{-1}$ wihle $U\sim \delta_\b^{-2}$. Therefore the leading contribution to the entropy at $\b_U$ is given by
\be
S=\frac{A}{4\ell_P^2}\lt[1+o(\l^{-1})+o(\delta_\b)\rt]-3\log \l,\nonumber
\ee
which reproduces the classical law $S=A_H/4\ell_P^2$ up to LQG corrections for infinitely many $\g = -\Z_+i+\eps$.


Before the analytic continuation, Chern-Simons level $k$ stands for the maximal area allowed at a single puncture (defect) on the horizon. The area of a single puncture should not be too large, otherwise it would break the macroscopic smoothness of the horizon. The situation is similar to the case of spinfoam LQG \cite{lowE}, where the spin should be cut-off by introducing quantum group or Chern-Simons theory \cite{QSF,CC}. The spin cut-off should not be too large, in order to preserve the macroscopic smoothness. 
Here $\l$ is assumed of the same scale as $k$. For example, if the spins are cut-off at the Grand Unification Scale, $k\ell_P^2$ or $\l\ell_P^2$ is the area scale of GUT, i.e. $k,\l\sim 10^6$. The Schwarzschild horizon area of the sun is $A_H\sim 10^6 m^2$. The maximal $\delta_\b\propto \sqrt{\l{\ell_P^2}/{A}}\sim 10^{-35}$ is a tiny LQG correction. This example also illustrates our approximation scheme $\delta_\b\ll 1/\l$ is natural.

As an analog of covariant LQG \cite{lowE}, $o(1/\l)$ or $o(1/k)$ are the quantum corrections relating to the large-$j$ expansion near the cut-off, while $o(\delta_\b)$ are high curvature UV corrections since $A$ relates to the curvature radius. The analysis here is valid in a semiclassical low energy regime $\ell_P^2\ll k\ell_P^2\ll A$. It is consistent with the proposal in \cite{GNP}.

Interestingly there exists a dual statistical system emerges from the partition function $Z_N$ by the above analysis, although its expression Eq.\ref{ZN} loses the obvious statistical interpretation as $\g=-i\eta=-i\eta_0+\eps$, $\eta_0\in\Z_+$. As $\b\to\b_U$ from $\b>\b_U$, the leading contribution to $Z_N$ in Eq.\Ref{star}, which is responsible for the leading energy and entropy, can be written as an integral up to prefactor that becomes logarithmic corrections in $\log\sz$, 
\be
Z_N\propto \frac{1}{N!}\int_{\R_+^N}\rmd^Ns\prod_{l=1}^N e^{2\pi \eta\zeta s_l -\b\eta \frac{\ell_P^2}{\ell}\zeta s_l},\ \ \ \zeta=\frac{\pi x}{\l|\chi|}>0\label{sint}
\ee
which interprets $Z_N$ as a statistical system with continuous energy spectrum $E=\eta_0\frac{\ell_P^2}{\ell}\zeta\sum_{l=1}^Ns_l$ ($s_l>0$) and degeneracy $d(\vec{s})=e^{2\pi \eta_0\zeta\sum_{l=1}^Ns_l}$. It implies that by analytic continuation, there exists a dual quantum theory of LQG with $\g= -i\Z$, which has a semiclassically continuous area spectrum $A=8\pi\eta_0\ell_P^2\zeta\sum_ls_l$ by $E=\frac{A}{8\pi\ell}$. The near-horizon quantum geometry is described in dual quantum theory by the number $N$ of punctures and a set of dual quantum areas $\{s_l\}_{l=1}^N$. Then importantly, the degeneracy of the dual quantum system is holographic, by 
\be
\log d(\vec{s})={\frac{A}{4\ell_P^2}},\label{logd}
\ee 
which shows that the maximal number of black hole microstates of a given near-horizon quantum geometry $\{s_l\}_{l=1}^N$ is given by the Bekenstein bound. 

In the case of Ashtekar variable with $\eta_0=1$, and if one takes $x=1,|\chi|=\frac{\pi}{\l}$ ($m=1$ in Eq.\Ref{f}), the degeneracy in the dual system Eq.\Ref{sint} reduces to $d(\vec{s})= e^{2\pi\sum_{l=1}^Ns_l}$, whose origin is exactly the factor $\prod_l\sin\frac{\pi d d_l}{k+2}$ in highest term $d=k+1$ in the Verlinde formula Eq.\Ref{verlinde}. In \cite{FGNP}, by complexifying the spins $j=is-\half$ and take $s,k$ to be large, $d=k+1$ term is picked up as the leading order, and the factor $\prod_l\sin\frac{\pi d d_l}{k+2}$ transforms into $e^{2\pi\sum_{l=1}^Ns_l}$. It has not been clarified in \cite{FGNP} if $e^{2\pi\sum_{l=1}^Ns_l}$ counts the quantum states of any system. However, from the above analysis, the result from \cite{FGNP} is justified as a state-counting in the dual quantum theory in the special case $\eta_0=1$. Furthermore the assumption of holographic degeneracy in \cite{GNP} also receives a justification here.

The dual statistical system Eq.\Ref{sint} or $\sz$ can be understood as $\int D g^{(2)}\exp[-(\b\frac{\ell_P^2}{\ell}-2\pi)\frac{A[g^{(2)}]}{8\pi\ell_P^2}]$. $g^{(2)}$ denotes a metric on the near-horizon 2-surface. It's consistent with an Euclidean path integral of Einstein gravity with a conical deficit angle $2\pi-\b\frac{\ell_P^2}{\ell}$ at the horizon \cite{GH,frodden}. It justifies the argument in \cite{GNP} which based on the assumption of holographic degenercy. It also suggests that there should be a derivation of Eq.\Ref{sint} from covariant LQG via semiclassical low energy approximation, given that covariant LQG reproduces Einstein gravity in the semiclassical low energy regime \cite{lowE,HZ}. Such a top-down approach to black hole thermodynamics is a research undergoing.

Finally, we remark that although the above derivation is for $\eta_0>0$, the generalization to $\eta_0<0$ ($k=-i\l,\ \l>0$ correspondingly) is straightforward, and only amounts to generalize the dual area spectrum by $A=8\pi|\eta_0\zeta|\ell_P^2\sum_ls_l$ and the holographic degeneracy by $\log d(\vec{s})=2\pi |\eta_0\zeta|\sum_{l=1}^Ns_l$. All the above results are valid to all $\eta_0\in \Z\setminus\{0\}$.


\section*{Acknowledgments}
 
The author thanks A. Perez for discussions and introducing me to his recent results. He also thanks M. Zhang for discussions. He gratefully acknowledges the CERN Winter School on Supergravity, Strings, and Gauge Theory 2014, and the 2rd EFI winter conference on Quantum Gravity, for the hospitality during this work. The research has received funding from the People Programme (Marie Curie Actions) of the European Union's 7th Framework Programme (FP7/2007-2013) under REA grant agreement No. 298786.


\begin{thebibliography}{20}

\bibitem{FGNP}E. Frodden, M. Geiller, K. Noui, A. Perez. [arXiv:1212.4060]

\bibitem{BH}J. Bardeen, B. Carter, S. Hawking. Comm. Math. Phys. 31 (1973) 161\\
J. Bekenstein. Phys. Rev. D7 (1973) 2333\\
S. Hawking. Comm. Math. Phys. 43 (1975) 199

\bibitem{bound}J. Bekenstein. Phys. Rev. D23 (1981) 287\\
R. Bousso. JHEP 9907 (1999) 004

\bibitem{holo}R. Bousso. Rev. Mod. Phys. 74 (2002) 825

\bibitem{rev}A. Ashtekar and J. Lewandowski. {Class. Quant. Grav.} {21} (2004) R53.\\
M. Han, W. Huang and Y. Ma. Int. J. Mod. Phys. D16 (2007) 1397-1474 [arXiv:gr-qc/0509064].

\bibitem{BHLQG}C. Rovelli, Phys. Rev. Lett. 77 (1996) 3288\\ 
A. Ashtekar, J. Baez, A. Corichi, K. Krasnov, Phys. Rev. Lett. 80 (1998) 904 \\
R. K. Kaul, P. Majumdar, Phys. Lett. B439 (1998) 267\\
A. Corichi, [arXiv:0901.1302 [gr-qc]].\\
M. Domagala and J. Lewandowski, Class. Quant. Grav. 21 (2004) 5233 

\bibitem{ENP}J. Engle, K. Noui, A. Perez. Phys. Rev. Lett. 105 (2010) 031302  

\bibitem{revisit}J. Engle, K. Noui, A. Perez, D. Pranzetti. JHEP 1105 (2011) 016 

\bibitem{GP}A. Ghosh, A. Perez. Phys. Rev. Lett. 107 (2011) 241301

\bibitem{FGP}E. Frodden, A. Ghosh, A. Perez. Phys. Rev. D 87 (2013) 121503

\bibitem{ash}A. Ashtekar. Phys. Rev. Lett. 57 (1986) 2244

\bibitem{P}D. Pranzetti. [arXiv:1305.6714]

\bibitem{GNP}A. Ghosh, K. Noui, A. Perez. [arXiv:1309.4563]

\bibitem{aldo}A. Riello. Phys. Rev. D 88 (2013) 024011 


\bibitem{bianchi}K. Krasnov, C. Rovelli. Class. Quant. Grav.26 (2009) 245009\\
E. Bianchi. Class.Quant.Grav.28 (2011) 114006


\bibitem{verlinde}E. Verlinde. Nucl. Phys. B 300 (1988) 360-376

\bibitem{BT}M. Blau and G. Thompson. Nucl.Phys. B408 (1993) 345-390

\bibitem{igamma}E. Frodden, M. Geiller, K. Noui, A. Perez. JHEP 05 (2013) 139\\
N. Bodendorfer. Phys. Lett. B726 (2013) 887 

\bibitem{seiberg}S Elitzur, G Moore, A Schwimmer, and N Seiberg. Nucl. Phys. B326 (1989) 108

\bibitem{lowE}M. Han. [arXiv:1308.4063]\\
M. Han. Phys. Rev. D 88 (2013) 044051 [arXiv:1304.5628]\\
M. Han. Class. Quantum Grav. 31 (2014) 015004 [arXiv:1304.5627]

\bibitem{QSF}M. Han. J. Math. Phys. 52 (2011) 072501 [arXiv:1012.4216]\\
W. Fairbairn, C. Meusburger. J. Math. Phys. 53 (2012) 022501

\bibitem{CC}M. Han. Phys. Rev. D 84 (2011) 064010 [arXiv:1105.2212]


\bibitem{GH}G. W. Gibbons, S. W. Hawking. Phys. Rev. D15 (1977) 2752\\
D. V. Fursaev, S. N. Solodukhin, Phys. Rev. D 52 (1995) 2133

\bibitem{frodden}E. Frodden, Ph.D. Thesis

\bibitem{HZ}M. Han, M. Zhang.  Class. Quantum Grav. 29 (2012) 165004 [arXiv:1109.0500]\\
M. Han, M. Zhang. Class. Quantum Grav. 30 (2013) 165012 [arXiv:1109.0499]\\
M. Han, T. Krajewski.  Class. Quantum Grav. 31 (2014) 015009 [arXiv:1304.5626]




\end{thebibliography}
\end{document}